\documentclass[12pt,preprint]{aastex}

\def\VEC#1{\mbox{\boldmath $#1$}}

\shorttitle{Energy Extraction by Magnetic Reconnection in Ergosphere}
\shortauthors{Koide \& Arai}

\begin{document}


\title{ Energy Extraction from a Rotating Black Hole by
Magnetic Reconnection in Ergosphere}


\author{Shinji Koide and Kenzo Arai}
\affil{Department of Physics, Kumamoto University, 
    2-39-1, Kurokami, Kumamoto 860-8555, Japan}
\email{koidesin@sci.kumamoto-u.ac.jp}




\begin{abstract}
We investigate mechanisms of energy extraction from a rotating black hole
in terms of negative energy-at-infinity.
In addition to the Penrose process through particle fission,
the Blandford-Znajek mechanism by magnetic tension,
and the magnetohydrodynamic Penrose process,
we examine energy extraction from a black hole caused 
by magnetic reconnection in the ergosphere. 
The reconnection redistributes
the angular momentum efficiently to yield the negative energy-at-infinity.
We derive a condition for the process to operate in a simple situation, 
where the plasma is incompressible and the magnetic energy 
is converted completely to the plasma kinetic energy locally.
Astrophysical situations of magnetic reconnection around the black holes
are also discussed.

\end{abstract}


\keywords{ Black hole physics, magnetohydrodynamics: MHD,
relativity, methods: analytical, galaxies: nuclei,
gamma rays: bursts, plasmas}



\section{Introduction}

Energy extraction from a rotating black hole interests us
not only as engines of relativistic jets from active galactic nuclei
(AGNs), micro-quasars ($\mu$QSOs), and gamma-ray bursts (GRBs) 
\cite{meier01}, but also
as fundamentals of black hole physics.
The horizon of the black hole is defined as the surface where no matter,
energy, and information pass through outwardly. 
On the other hand, reducible energy from the rotating black hole is given by
\begin{equation}
E_{\rm rot} = \left [ 1- \sqrt{\frac{1}{2} \left (1+\sqrt{1-a^2} \right )} 
\right ] Mc^2  ,
\end{equation}
where $M$ is the mass, $a$ is the rotation parameter of the black hole, 
and $c$ is the speed of light \cite{misner70}.
It corresponds to the rotational energy that can be
extracted principally. 

Several kinds of distinct mechanisms have been proposed for the extraction
of the rotational energy: e.g. Penrose process, super-radiant scattering,
Blandford-Znajek mechanism, magnetohydrodynamic (MHD) Penrose process, and
modified Hawking process  
\cite{penrose69,press72,ruffini75,blandford77,hirotani92,putten00}.
Here we mention three kinds of mechanism of the black hole energy extraction
among them: the Penrose process, the Blandford-Znajek mechanism,
and the MHD Penrose process, while the super-radiant scattering and the
modified Hawking process may be related with the high energy phenomena
such as origins of gamma-ray bursts and the ultra-high energy cosmic rays
\cite{putten99,putten00,pierre07}.
The Penrose process involves production of particles with
negative energy-at-infinity via strong fission or particle interaction 
in the ergosphere \cite{penrose69}. 
It needs drastic redistribution of angular momentum 
to produce the negative energy-at-infinity.
%
Although this process clearly shows a possibility of energy extraction from
the black hole, it is improbable as engines of 
astrophysical jets, because of poor collimation of particles 
and poor event rate. That is, the Penrose process accelerates
the particles toward the equatorial plane, not toward the axis direction,
and needs the azimuthal relativistic fission at not so wide region
in the ergosphere.
Blandford \& Znajek (1977) proposed a mechanism of energy extraction from a 
rotating black hole through force-free magnetic field.
Their analytic steady-state solutions show the direct energy radiation
from the horizon, which looks contradictory to the definition of the black 
hole horizon \cite{punsly90}. As we discuss in the next section, 
this mechanism also utilizes the negative energy-at-infinity.
In this case, however, the negative energy-at-infinity is sustained 
not by the particle or matter, but by the electromagnetic field. 
Angular momentum of the electromagnetic field is redistributed by
the magnetic tension through almost mass-less plasma
to produce the negative energy-at-infinity of the field.
%
The magnetic tension may also redistribute the angular momentum
of the plasma to yield the negative energy-at-infinity of the plasma,
when the plasma has non-zero mass density \cite{hirotani92}.
It is called the MHD Penrose process.
This energy extraction was confirmed by the numerical simulations based on the general
relativistic magnetohydrodynamics (GRMHD) \cite{koide02,koide03}.

%
%
It is noted that magnetic reconnection also redistributes 
angular momentum of the plasma to form the negative energy-at-infinity
because it produces a pair of fast outflows with the opposite directions
from the reconnection region.
Then the rotational energy of the black hole can be extracted through
the induced negative energy-at-infinity of the plasma.
In the present paper, we derive a condition for the process to operate
in a simple situation for the incompressible plasma, 
where all the magnetic energy is converted to the plasma kinetic energy.

In \S \ref{oldmecha}, we review the mechanisms of energy extraction 
from the rotating black hole in terms of the negative energy-at-infinity.
In \S \ref{newmecha}, we examine the operation condition of the energy
extraction from the black hole
induced by magnetic reconnection in the ergosphere using a simple model. 
In \S \ref{discussion}, we discuss astrophysical situations
where the magnetic reconnection happens in the ergosphere.

\section{Penrose process, Blandford-Znajek mechanism, 
and MHD Penrose process} \label{oldmecha}

We use the Boyer-Lindquist coordinates
$(ct, r, \theta, \phi)$ to describe the space-time
around a rotating black hole.
The line element of the axisymmetric, stationary space-time
around the black hole is written by
\begin{equation}
ds^2 = -\alpha^2 (cdt)^2 + \sum_{i=1}^3 
h_i^2 \left ( dx^i - \omega_i dt \right )^2
= -\alpha^2 (cdt)^2 + \sum_{i=1}^3 
\left ( h_i dx^i - c \alpha \beta_i dt \right )^2   ,
\label{kerrmetric}
\end{equation}
where $h_i$ is the scale factor of the coordinate
$x^i$, $\omega_i$ is the angular velocity describing
a frame-dragging effect, $\alpha$ is the lapse function, and
${\beta}_i= h_i \omega_i/(c \alpha)$ is the shift vector.
For the Kerr metric \cite{misner70}, we have
\begin{equation}
\alpha = \root \of {\frac{\Delta \Sigma}{A}}, \verb! !
h_1 =  \root \of {\frac{\Sigma}{\Delta}}, \verb! !
h_2 =  \root \of {\Sigma}, \verb! !
h_3 =  \root \of {\frac{A}{\Sigma}} \sin \theta, \verb! !
\omega_1=\omega_2=0, \verb! !
\omega_3=\frac{2cr_{\rm g}^2 a r}{A},
\end{equation}
where $\Delta = r^2-2 r_{\rm g} r + (ar_{\rm g})^2$,
$\Sigma = r^2 + (ar_{\rm g})^2 \cos^2 \theta$,
$A=\{ r^2 + (ar_{\rm g})^2 \}^2 - \Delta (ar_{\rm g})^2 \sin^2 \theta$, and
$r_{\rm g} = GM/c^2$ is the gravitational radius of the 
black hole.

This metric has translational symmetry with respect to $t$ and
$\phi$, so that we obtain the conservation law
\begin{equation}
\frac{1}{\root \of {- g}}
\frac{\partial}{\partial x^\mu}
\left ( \root \of {- g} T^{\mu\nu} \xi _\nu
\right ) =0  ,
\label{killingconserv}
\end{equation}
where $g = {\rm Det}(g_{\mu\nu})= -(\alpha h_1 h_2 h_3)^2$ is the determinant of 
the metric tensor, 
$T^{\mu\nu}$ is the energy-momentum tensor, and
$\xi^\nu$ is the Killing vector.
When we adopt one component approximation of the plasma, we have
\begin{equation}
T^{\mu\nu} = p g^{\mu\nu} + {\mathfrak h} U^\mu U^\nu + F^\mu _\sigma F^{\sigma \nu}
-\frac{1}{4} g^{\mu\nu}F^{\rho\sigma} F_{\rho \sigma},
\label{enemomenttensor}
\end{equation}
where $p$ is the proper pressure,
${\mathfrak h} = e_{\rm int} + p$ 
is the enthalpy density,
$U^\mu$ is the four-velocity, and $F_{\mu\nu}$ is the electromagnetic
field-strength tensor \cite{koide99}.
The thermal energy density is given by $e_{\rm int} = p/(\Gamma-1) + \rho c^2$
for adiabatic plasma,
where $\Gamma$ is the adiabatic index
and $\rho$ is the proper mass density.

When we consider the Killing vectors $\xi^\nu = (-1, 0, 0, 0)$ and (0, 0, 0, 1), 
we get the energy and angular momentum conservation laws,
\begin{eqnarray}
\frac{\partial e^\infty}{\partial t} 
& = & - \frac{1}{h_1 h_2 h_3} \sum_i 
\frac{\partial}{\partial x^i}
\left ( h_1 h_2 h_3 S^i\right )  ,  \\
\frac{\partial l}{\partial t} 
& = & - \frac{1}{h_1 h_2 h_3} \sum_i 
\frac{\partial}{\partial x^i}
\left ( h_1 h_2 h_3 M^i \right )  ,
\end{eqnarray}
where $e^\infty = - \alpha g_{\nu 0} T^{\nu 0}$ is called 
``energy-at-infinity" density,
which corresponds to the total energy density of the plasma and field, 
$S^i = - c \alpha g_{\nu 0} T^{i \nu}$ is the energy flux density,
%
$l = \alpha g_{3\nu} T^{3\nu}/c$ is the angular momentum density,
and $M^i  =  \alpha h_i T^{i\nu} g_{\nu 3}$ is the angular
momentum flux density.

When we introduce the local frame called the ``zero angular momentum observer"
(ZAMO) frame, we have $d\hat{t} = \alpha dt$, $d\hat{x}^i=h_i(dx^i-\omega_i dt)$.
Because this is the local Minkowski space-time:
$ds^2 = - (cd\hat{t})^2 + \sum_{i=1}^3 (d\hat{x}^i)^2 
= \eta_{\mu\nu} d\hat{x}^\mu d\hat{x}^\nu$,
the variables observed in the frame are intuitive.
For example, the velocity $\hat{v}^i$, the Lorentz factor
$\gamma = \hat{U}^0 = \left [ 1 - \sum_{i=1}^3 (\hat{v}^i/c)^2
\right ]^{-1/2}$, and the four-velocity 
$\hat{U}^i = h_i U^i - c \alpha \beta^i U^0$ $(i=1,2,3)$ have the relation,
$\hat{U}^i = \gamma \hat{v}^i $.
Hereafter we denote the variables observed in the ZAMO frame with the hat, 
$\wedge$.
From equation (\ref{enemomenttensor}), 
we obtain
\begin{eqnarray}
e^\infty &=& \alpha e + \sum_i \omega_i h_i \hat{P}^i =
\alpha e + \omega_3 l ,
\label{eatinfden}  \\
l        &=& h_3 \hat{P}^3 .
\label{angmomden}
\end{eqnarray}
Here $e = {\mathfrak h} \gamma^2 -p + \left (
\hat{B}^2 + \hat{E}^2/c^2 \right )/2$ is the total energy
density and $\hat{P}^i = \{ {\mathfrak h}\gamma^2 \hat{v}^i 
+ ( \hat{\VEC{E}} \times \hat{\VEC{B}})^i  \}/c^2$ 
is the $i$-th component of the momentum density,
where $\hat{B}^i = \epsilon^{ijk} \hat{F}_{jk}/2$ and
$\hat{E}^i = c \eta^{ij} \hat{F}_{j0} = c \hat{F}^i_0
= c \hat{F}_{i0}$ ($i=1,2,3)$ are the magnetic flux
density and the electric field, respectively ($\epsilon^{\mu\nu\lambda}$
is the Levi-Civita tensor).
We normalize the field strength tensor $F_{\mu\nu}$ so that
$|\hat{\VEC{B}}|^2/2$ and $|\hat{\VEC{E}}|^2/(2c^2)$ give the magnetic and
electric energy densities, respectively. For example, the magnetic field
measured in the SI unit divided by the square root of the magnetic permeability
$\mu_0$ is $\hat{\VEC{B}}$ and the electric field measured in the SI unit times
the square root of permittivity of vacuum $\epsilon_0$ is $\hat{\VEC{E}}$.
Equations (\ref{eatinfden}) and (\ref{angmomden}) can be separated into the
hydrodynamic and electromagnetic components:
$e^\infty = e^\infty_{\rm hyd} + e^\infty_{\rm EM}$ and
$l = l_{\rm hyd} + l_{\rm EM}$, where
\begin{eqnarray}
e^\infty_{\rm hyd} &=& \alpha \hat{e}_{\rm hyd} 
+ \sum_i \omega_i h_i \frac{\mathfrak h}{c^2} \gamma^2 \hat{v}^i =
\alpha \hat{e}_{\rm hyd} + \omega_3 l_{\rm hyd}  , 
\label{eatinfhyd} \\
e^\infty_{\rm EM}  &=& \alpha \hat{e}_{\rm EM} 
+ \sum_i \omega_i h_i \frac{1}{c^2} \left ( \hat{\VEC{E}} \times \hat{\VEC{B}}  
\right )_i =
\alpha \hat{e}_{\rm EM} + \omega_3 l_{\rm EM}  , 
\label{eatinfem}  \\
l_{\rm hyd} &=& h_3 \frac{{\mathfrak h}}{c^2} \gamma^2 \hat{v}^3  , 
\label{angmomhyd}  \\
l_{\rm EM} &=& h_3 \frac{1}{c^2} \left ({\hat{\VEC{E}}} \times \hat{{\VEC{B}}}
\right )_3  .
\label{angmomem}
\end{eqnarray}
Here, $\hat{e}_{\rm hyd} = {\mathfrak h} \gamma^2 -p$ and
$\hat{e}_{\rm EM} = ( \hat{B}^2 + \hat{E}^2/c^2 )/2$
are the hydrodynamic and electromagnetic energy densities observed by the ZAMO frame,
respectively.

\subsection{Penrose process}

When we consider a particle with rest mass $m$ at $\VEC{r}_{\rm P}(t)$ 
in the absence of electromagnetic field, {\it i.e.}
$\rho=\frac{m}{\gamma} \delta^3({\bf r} - \VEC{r}_{\rm P}(t))$,
$p=0$, and ${\VEC{B}}={\VEC{E}}={\bf 0}$, then 
the energy-at-infinity and the angular momentum of the particle are given
from equations (\ref{eatinfden}) and (\ref{angmomden}) as, 
\begin{eqnarray}
E^\infty &=& \int_\Omega e^\infty dV =
\alpha \gamma m c^2+ \omega_3 L, \label{eatinf}  \\
L &=& \int_\Omega l dV = h_3 m \gamma \hat{v}^3, \label{angmoment}
\end{eqnarray}
where $\Omega$ is the whole volume of the space.
Both the energy-at-infinity and angular momentum of the particle 
conserve when it travels alone. 

Equations (\ref{eatinf}) and (\ref{angmoment}) yield
the energy-at-infinity of the particle as
\begin{equation}
E^\infty = \alpha \gamma m c^2 \left ( 1 + \beta_3 \frac{\hat{v}^3}{c}
\right )  .
\end{equation}
If $\beta_3 \hat{v}^3/c < -1$, the energy-at-infinity of the particle
becomes negative. This condition can be satisfied only in the ergosphere
($\beta_3 > 1$). Using the relation 
$\alpha^2 \left \{ 1- \sum_i (\beta_i)^2 \right \} = -g_{00}$,
we have the well-known definition of the ergosphere: $g_{00} \ge 0$.

When we consider the particle fission in the ergosphere, A $\longrightarrow$ B + C,
the conservation laws of the energy-at-infinity and the angular momentum are 
\begin{eqnarray}
E^\infty_{\rm A} &=& E^\infty_{\rm B}+E^\infty_{\rm C} , \\
L_{\rm A} &=& L_{\rm B} + L_{\rm C}  ,\\
E^\infty_I &=& \alpha \gamma_I m_I c^2+ \omega_3 L_I \verb!    !
(I={\rm A}, {\rm B}, {\rm C}) .
\end{eqnarray}
If the fission is so strong that satisfies
$L_{\rm B} = h_3 m_{\rm B} \gamma_{\rm B} \hat{v}^3_{\rm B} 
< - ch_3 m_{\rm B}/\sqrt{(\beta_3)^2-1}$,
then we get the negative energy-at-infinity $E^\infty_{\rm B} < 0$ and
$E^\infty_{\rm A} < E^\infty_{\rm C}$.
The particle C escapes to infinity and the particle B
is swallowed by the black hole to reduce the black hole mass. 
Eventually, the rotational energy of the black hole is extracted.

\subsection{Blandford-Znajek mechanism}

In the Blandford-Znajek mechanism,
the energy propagates
outwardly from the black hole horizon when the angular velocity
of the black hole horizon $\Omega_{\rm H}$ is larger than that
of the magnetic field lines $\Omega_{\rm F}$: 
$\Omega_{\rm H} > \Omega_{\rm F}$.
Although this statement looks inconsistent to the definition of the horizon, 
where any energy, matter, and information never transform outwardly
\cite{punsly90}, it can be understood as the transportation of the 
negative electromagnetic energy-at-infinity of magnetic fields 
into the black hole \cite{koide03}.
Here we show the electromagnetic energy-at-infinity
becomes negative when $\Omega_{\rm H}>\Omega_{\rm F}$.

From equations (\ref{eatinfem}) and (\ref{angmomem}) and the definition of 
$\hat{e}_{\rm EM}$ and $\hat{P}^i$,
the electromagnetic energy-at-infinity $e^\infty_{EM}$
is written as
\begin{equation}
e^\infty_{\rm EM} = \frac{\alpha}{2} \left ( \hat{B}^2+\frac{\hat{E}^2}{c^2}
\right ) + \alpha \VEC{\beta} \cdot \left ( \hat{\VEC{E}} \times \hat{\VEC{B}}
\right ) .
\label{engatinfem}
\end{equation}
When we assume steady-state of the electromagnetic field in
the force-free condition, the electric field observed by the ZAMO frame is given by
\begin{equation}
\hat{\VEC{E}} = - \frac{h_3}{\alpha} \left ( 
\Omega_{\rm F} - \omega_3 \right )
{\bf e}_\phi \times \hat{\VEC{B}}^{\rm P}   ,
\label{elefldfrcfre}
\end{equation}
where ${\bf e}_\phi$ is the unit vector parallel to the azimuthal
coordinate,
$\hat{\VEC{B}}^{\rm p} = \hat{\VEC{B}} - \hat{B}_\phi {\bf e}_\phi$
is the poloidal magnetic field,
and $\Omega_{\rm F}$ is a constant along the magnetic flux tube.
If we use the velocity of the magnetic flux tubes
observed in the ZAMO frame, $\hat{\VEC{v}}_{\rm F} = (h_3/\alpha)
(\Omega_{\rm F} - \omega_3 ) {\bf e}_\phi$, equation (\ref{elefldfrcfre}) can
be written as a intuitive equation,
$ \hat{\VEC{E}} = - \hat{\VEC{v}}_{\rm F} \times \hat{\VEC{B}}^{\rm P}   .$
Inserting equations (\ref{elefldfrcfre}) into (\ref{engatinfem}), we have
the electromagnetic energy-at-infinity,
\begin{equation}
e^\infty_{\rm EM} = \alpha \frac{(\hat{B}^{\rm P})^2}{2}
\left ( \frac{h_\phi}{c\alpha} \right )^2
\left [ \Omega_{\rm F}^2 - (\omega_3)^2
\left \{  1 - \frac{1}{(\beta_3)^2} 
- \left ( \frac{\hat{B}_\phi}{\beta_3 \hat{B}^{\rm P}} \right )^2
\right \} \right ] .
\label{eatinfff}
\end{equation}
We consider the electromagnetic energy-at-infinity very near the horizon.
We adopt the boundary condition of the electromagnetic field at the horizon,
\begin{equation}
\frac{\hat{B}_\phi}{\hat{B}^{\rm P}} 
= \frac{\hat{v}^\phi_{\rm F}(r \rightarrow r_{\rm H})}{c} ,
\label{bundcondathoriz}
\end{equation}
where $r_{\rm H}$ is the radius of the horizon.
The condition (\ref{bundcondathoriz}) 
is intuitive when we assume the Alfven velocity 
is the light speed $c$ in the force-free magnetic field.
It is also identical to the condition used 
by Blandford \& Znajek (1977), while the original expression
is rather complex.
After some manipulations with equations 
(\ref{eatinfff}) and (\ref{bundcondathoriz}), we obtain at the horizon
\begin{equation}
\alpha e^\infty_{\rm EM} = \left ( \frac{h_3}{c} \right )^2
(\hat{B}^{\rm P})^2  \Omega_{\rm F} (\Omega_{\rm F} - \omega_3)  .
\end{equation}
Here we have used $\alpha \rightarrow 0$ very near the horizon.
The electromagnetic energy-at-infinity is negative only when
$0 < \Omega_{\rm F}< \omega_3$, which 
is identical to the switch-on condition of the
Blandford-Znajek mechanism.
The coincidence indicates that the Blandford-Znajek
mechanism utilizes the negative electromagnetic energy-at-infinity.
This conclusion is generally applicable to the force-free field of 
any spinning black hole as far as equation (\ref{bundcondathoriz})
is valid, while the original analytical model of Blandford \& Znajek 
(1977) is restricted to a slowly spinning black hole.
In the force-free condition, the Alfven velocity becomes the speed of
light, and then the Alfven surface is located at the horizon.
Then the region of the negative energy-at-infinity can be connected
with the region outside of the ergosphere through the magnetic field
causally.

\subsection{MHD Penrose process}

The MHD Penrose process is the mechanism of energy extraction 
from a black hole through 
the negative energy-at-infinity of plasma induced by the magnetic
tension \cite{hirotani92}.
It has been confirmed by Koide et al. (2002) and Koide (2003) 
using GRMHD numerical simulations. 
It is defined as the energy extraction mechanism with the negative
energy-at-infinity of plasma, which is induced by the magnetic tension, 
while in the Blandford-Znajek mechanism,
the negative electromagnetic energy-at-infinity plays a important role.
For a rapidly rotating black hole ($a=0.99995$),
the magnetic flux tubes of the strong magnetic field which cross the ergosphere
are twisted due to the frame-dragging effect.
The angular momentum of plasma in the ergosphere is opposite to
that of the rotating black hole and its magnitude is large to make
$e^\infty_{\rm hyd}$ negative in equation (\ref{eatinfhyd}).
The twist of the magnetic flux tubes propagates outwardly. 
The Poynting flux indicates that the electromagnetic energy is radiated from the
ergosphere (see figure 4 of Koide et al. 2003). 
At the foot point of the energy radiation from the ergosphere,
the hydrodynamic energy-at-infinity 
$e^\infty_{\rm hyd} = \alpha \gamma \hat{e}_{\rm hyd} + \omega_3 l_{\rm hyd}$
decreases rapidly and becomes negative quickly.
The negative energy-at-infinity is mainly composed of that of the plasma.
To realize the negative energy-at-infinity of the plasma,
redistribution of the angular momentum of the plasma,
$l_{\rm hyd} = h_3 {\mathfrak h} \gamma^2 \hat{v}_\phi$, is demanded.
The angular momentum of the plasma is mainly redistributed by the
magnetic tension.

\section{Magnetic reconnection in ergosphere} \label{newmecha}

\begin{figure}
\epsscale{1.00}
\plotone{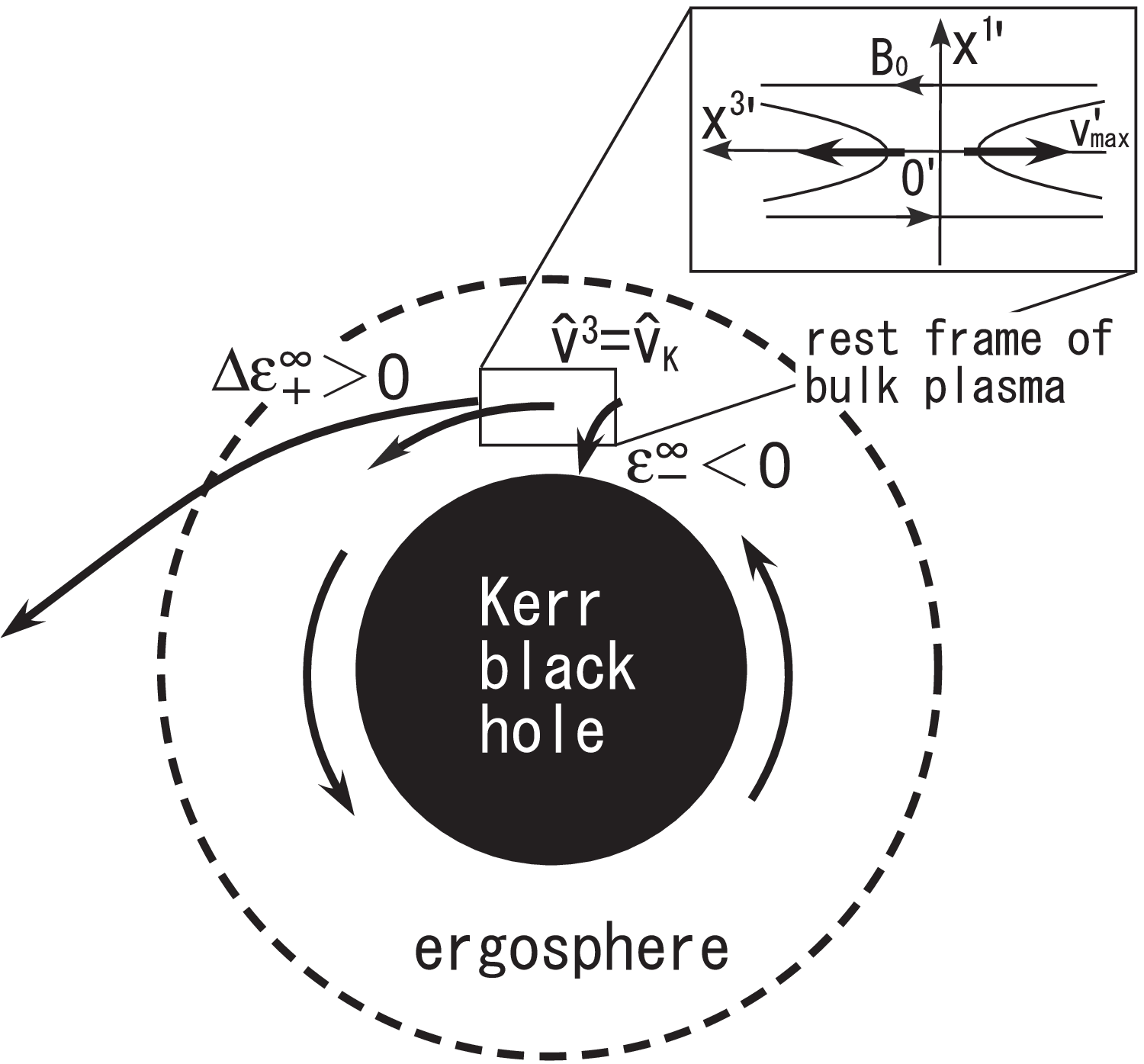}
\caption{
Schematic picture of the extraction of the black hole energy 
through the magnetic reconnection in the bulk plasma which rotates
circularly on the equatorial plane of the rotating black hole.
The phenomena are indicated in the Boyer-Lindquist coordinates.
The coordinates ${\rm O}'-x^{1\prime}x^{2\prime}x^{3\prime}$ 
in the inserted box are the rest frame of the bulk plasma.
\label{zentaipontie}}
\end{figure}

We investigate energy extraction through negative energy-at-infinity
induced by magnetic reconnection in an ergosphere around
a Kerr black hole.
For simplicity, we consider the magnetic reconnection in the bulk 
plasma rotating around the black hole circularly at the equatorial
plane (Fig. \ref{zentaipontie}).
To sustain the circular orbit, the plasma rotates with the
Kepler velocity $v_{\rm K}$ or it is supported by external force,
like magnetic force. 
Here the Kepler velocity is given by
\begin{equation}
\hat{v}_{\rm K} = \frac{cA\left [ \pm \sqrt{r_{\rm g}/r} 
- a r_{\rm g}^2/r^2 \right ]}{\sqrt{\Delta}(r^3-r_{\rm g}^3a^2)} 
-c\beta^3.
\end{equation}
The plus (minus) sign corresponds to the co-rotating 
(counter-rotating) circular orbit case.
Throughout this paper, we use the co-rotating Kepler velocity.
We assume the initial anti-parallel magnetic field directs
to the azimuthal direction in the bulk plasma, and
the pair plasma outflows moving toward the opposite directions each other
caused by the magnetic reconnection are ejected
in the azimuthal direction. 
It is also assumed that the plasma acceleration through 
the magnetic reconnection is localized in the very small region 
compared to the size of the black hole ergosphere, and 
the magnetic field outside of the plasma acceleration region is so weak that  
the plasma flow accelerated by the magnetic reconnection is not influenced
by the large-scale magnetic field around the black hole.
If one of the pair plasma flows in the opposite direction of the black hole
rotation has negative energy-at-infinity
and the other in the same direction of the black hole rotation 
has the energy-at-infinity higher than the rest mass energy
(including the thermal energy) (Fig. \ref{zentaipontie}), 
the black hole rotational
energy will be extracted just like the Penrose process.

\begin{figure}
\epsscale{.55}
\plotone{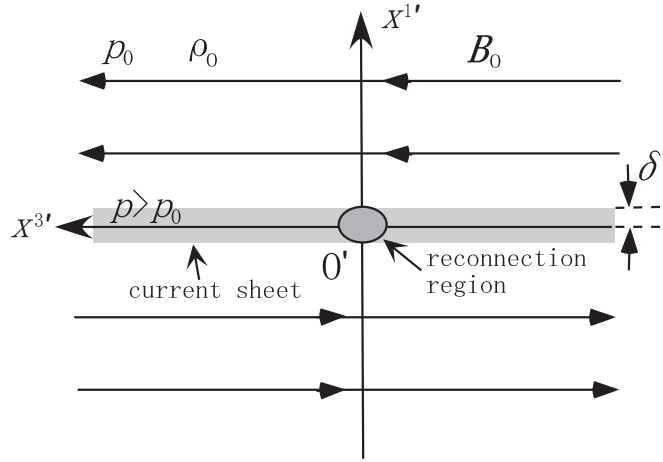}
\caption{Initial magnetic configuration around the reconnection region in
the local rest frame of the bulk plasma. 
The area in the dark gray ellipse at the origin corresponds to the magnetic
reconnection region and the gray layer along the $x^{3'}$ shows
the current sheet.
\label{magconfker}}
\end{figure}

\begin{figure}
\epsscale{.55}
\plotone{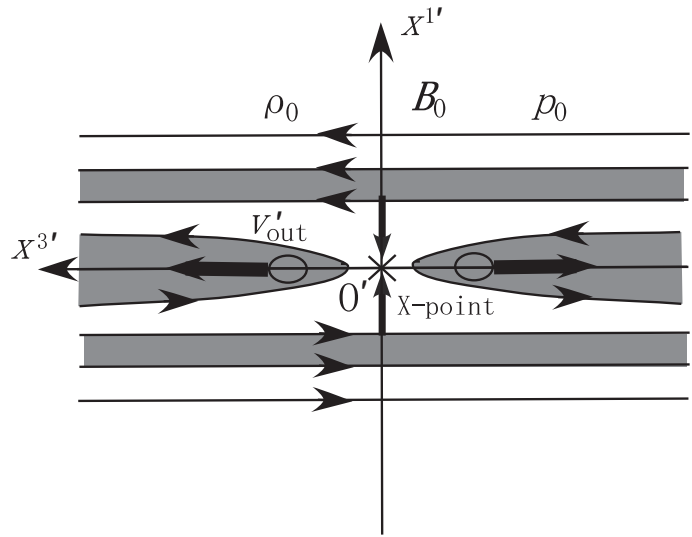}
\caption{
Energy convergence from the magnetic energy to the kinetic
energy through the magnetic reconnection
in the rest frame of the bulk plasma.
\label{magconfmin}}
\end{figure}

We have to investigate two conditions, {\it i.e.} the condition for
the formation of the negative energy-at-infinity and the condition for
escaping to infinity.
Before we move on to the conditions, we here mention about the elementary
process of the relativistic magnetic reconnection in the locally uniform,
small-scale plasma rotating circularly around the black hole.
%
To investigate the magnetic reconnection in the small scale, 
we introduce the local rest frame 
$(ct',x^{1\prime},x^{2\prime},x^{3\prime})$ 
of the bulk plasma which rotates
with the azimuthal velocity $\hat{v}^3= c\beta_0$
at the circular orbit on the
equatorial plane $r=r_0<r_{\rm S}$, $\theta=\pi/2$.
We set the frame $(ct',x^{1\prime},x^{2\prime},x^{3\prime})$
so that the direction of $x^{1\prime}$ coordinate is parallel to the
radial direction $x^1=r$ and the direction of $x^{3\prime}$ is
parallel to the azimuthal direction $x^3=\phi$ (Fig. \ref{zentaipontie}).
Hereafter we note the variables observed by the rest frame of the 
bulk plasma with the prime, ``$\prime$".
First, we consider the magnetic reconnection in the rest frame locally and
neglect the tidal force and Coriolis' force for simplicity.
We regard that the rest frame rotating with the Kepler velocity
is in a gravity-free state and
%
thus we can consider the magnetic reconnection 
in the framework of special relativistic MHD.
The initial condition is set to that of the Harris model 
where the anti-parallel magnetic field and the plasma are
in equilibrium:
\begin{eqnarray}
B^{3\prime} &=& B_0 \tanh (x^{1\prime}/\delta), \verb! ! B^{1\prime}= B^{2\prime}= 0, \\
p &=& \frac{B_0^2}{2\cosh ^2 (x^{1\prime}/\delta)}+ p_0, \\
\rho &=& \rho_0, \\
v^{1\prime}  &=&  0 ,\hspace{0.5cm} v^{2\prime}=0,
\hspace{0.5cm} v^{3\prime} = 0,
\end{eqnarray}
where $B_0$ is the typical magnetic field strength and
$2\delta$ is the thickness of the current layer
(see Fig. \ref{magconfker}).
The electric resistivity is assumed to be zero except for 
the narrow reconnection region. 
We assume the system is symmetric with respect to the $x^{2\prime}$-direction,
and resistivity is given by
\begin{equation}
\eta = \eta_0 f(x^{1\prime},x^{3\prime})  ,
\end{equation}
where $\eta_0$ is a positive constant and $f(x^{1\prime},x^{3\prime})$ is 
the profile of the resistivity,
which is finite around $x^{1\prime}=0$, $x^{3\prime}=0$ but zero outside
the reconnection region.
The magnetic flux tubes reconnected at the resistive region
accelerate the plasma through the magnetic tension,
and the magnetic energy in the flux tubes is converted to the
kinetic energy of the plasma. As shown in Fig. \ref{magconfmin},
the magnetic flux tube and the
plasma run away and fresh tubes and the plasma
are supplied to the reconnection region from outside of the current layer
successively.

\begin{figure}
\epsscale{.8}
\plotone{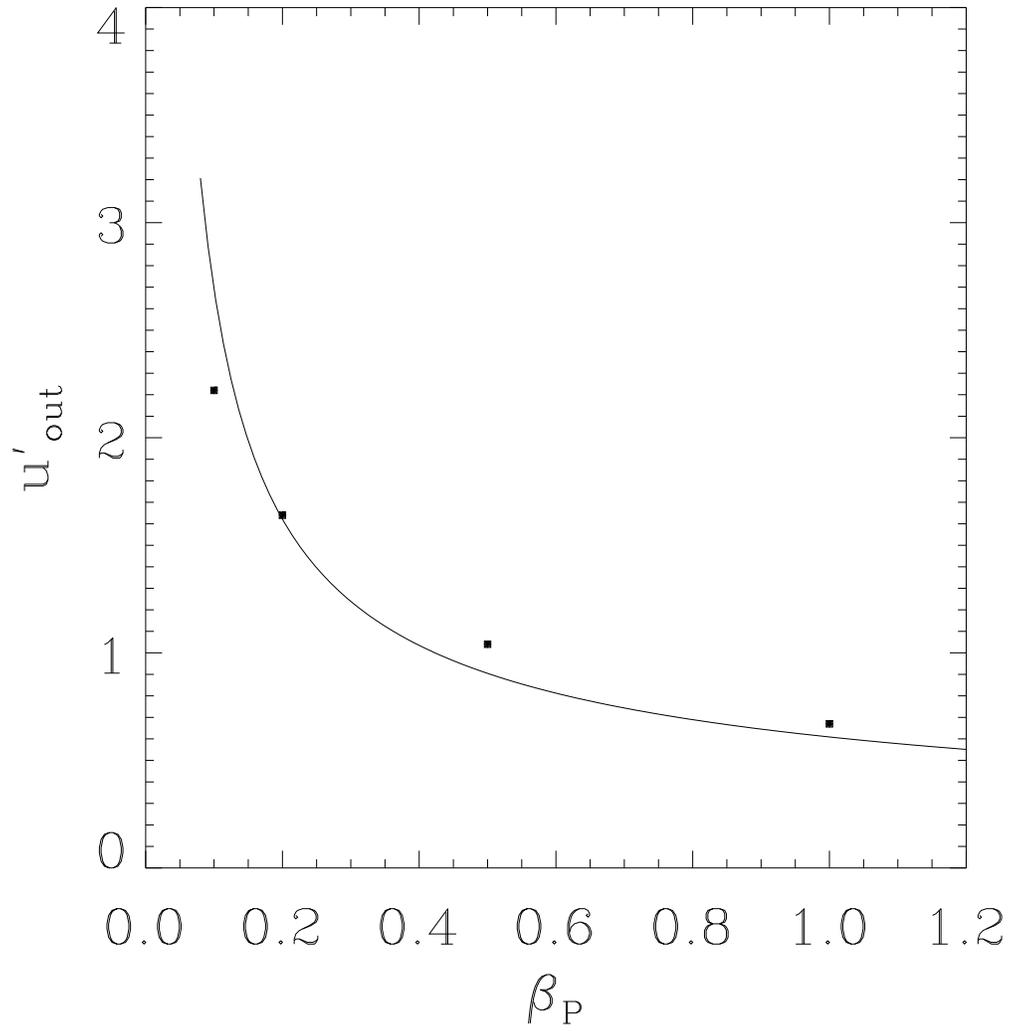}
\caption{
Comparison between the numerical result of the four-velocity
of the plasma outflow $u'_{\rm out}$ through the magnetic 
reconnection (full squares; Watanabe et al. 2006) 
and the simple expression (\ref{fvlmaxrec}) (solid line) as a function
of the plasma beta, $\beta_{\rm P}$.
\label{compare}}
\end{figure}

The outflow velocity of the accelerated plasma through the
magnetic reconnection, $v'_{\rm out}$, is estimated
by the velocity $v'_{\rm max}$, where whole magnetic energy
is converted to the kinetic energy.
If the magnetic energy is completely converted to the kinetic energy of
the plasma particles which are initially at rest, 
the magnetic energy per plasma particle is $B_0^2/(2n_0)$,
where $n_0$ is the plasma particle number density. 
Using the approximation that the plasma element is treated
as incompressible gas covered by very thin, light, adiabatic skin
with the thermal energy $U$ and 
the enthalpy $H$, using equation (\ref{apptotengspe}) in Appendix A, 
we can write the energy 
conservation equation with respect to the particle as,
\begin{equation}
\gamma'_{\rm max} H - \frac{(\Gamma-1) U}{\gamma'_{\rm max}}
= \frac{B_0^2}{2n_0} + H -(\Gamma-1) U,
\label{energybalonep}
\end{equation}
where $\gamma'_{\rm max}$ is the Lorentz factor of the plasma particle 
after complete release of the magnetic field energy
observed in the bulk plasma rest frame.
We get the maximum Lorentz factor $\gamma'_{\rm max}$
from equation (\ref{energybalonep}) as,
\begin{equation}
\gamma'_{\rm max} = \frac{1}{4} \left [ 
u_{\rm A}^2 + 2 (1-\varpi) + \sqrt{4D} \right ]  ,
\label{lofmaxrec} 
\end{equation}
where $u_{\rm A}^2 = B_0^2/h_0$, $\varpi = p_0/h_0$, 
$4D=[u_{\rm A}^2+ 4(1-\varpi)]^2+16 \varpi$, 
$p_0=(\Gamma-1)Un_0$ is the pressure, and $h_0=n_0 H$ is the enthalpy density
of the plasma.
Obviously, the terminal velocity of the plasma outflow through the 
magnetic reconnection is smaller than the maximum velocity $v'_{\rm max}$,
since all the magnetic energy is not always converted to the kinetic energy 
because of the Joule heating in the reconnection region and finite length acceleration.
Figure \ref{compare} shows the four-velocity 
$u'_{\rm out} = \gamma'_{\rm out} v'_{\rm out}$
of the plasma outflow caused by the magnetic reconnection
against the plasma beta $\beta_{\rm P}=2p_0/B_0^2$ in the case of $\Gamma=4/3$ 
and $B_0^2=\rho_0c^2$.
The solid line denotes the predicted values from equation (\ref{lofmaxrec}),
\begin{equation}
u'_{\rm max} = \gamma'_{\rm max} \frac{v'_{\rm max}}{c}
=\frac{1}{2} \left [
\left ( 1 + \frac{u_{\rm A}^2}{2} -\varpi \right )^2
+ 2(\varpi -1) + \left ( 1 + \frac{u_{\rm A}^2}{2} - \varpi \right ) \sqrt{D}
\right ]  ,
\label{fvlmaxrec}
\end{equation}
where 
\begin{equation}
u_{\rm A} = \left [ 
\frac{\rho_0c^2}{B_0^2} + \frac{\Gamma}{2(\Gamma-1)} \beta_{\rm P}
\right ]^{-1/2}, \hspace{0.5cm}
\varpi = \frac{\beta_{\rm P}}{2} \left [ 
\frac{\rho_0c^2}{B_0^2} + \frac{\Gamma}{2(\Gamma-1)} \beta_{\rm P}
\right ]^{-1}  .
\end{equation}
The full squares are from the numerical calculations (Watanabe et al. 2006).
It is found that our values of $u'_{\rm max}$ are 
in good agreement with the result of numerical simulations.
In the high plasma beta region, the values of $u'_{\rm max}$ from
equation (\ref{fvlmaxrec}) are slightly
smaller than those of the numerical result.
This discrepancy, {\it i.e.} $u'_{\rm max} < u'_{\rm out}$,
should come from the release of the thermal energy to
the plasma kinetic energy.
Therefore, we take the maximum
velocity $v'_{\rm max}$ as the plasma outflow velocity
induced by the magnetic reconnection hereafter.

Now we return to consider the conditions of the negative energy-at-infinity
formation and escaping to infinity of the pair outflows
caused by the magnetic reconnection.
From equations (\ref{appengatinfbal}),
the hydrodynamic energy-at-infinity per enthalpy of the plasma ejected
through the magnetic reconnection into the $\pm x^{3\prime}$ (azimuthal) 
direction with the Lorentz factor $\gamma'_{\rm max}$ 
is given by
\begin{eqnarray}
& &\epsilon_\pm^\infty (u_{\rm A}, \hat{\beta}_0, \varpi, \alpha, \beta_3) 
\equiv \frac{e^\infty_{{\rm hyd},\pm}}{h_0}
= \frac{E^\infty_\pm}{H} \nonumber \\
&=& \alpha \hat{\gamma}_0 \left [
(1+\beta_3 \hat{\beta}_0) \gamma'_{\rm max} \pm (\hat{\beta}_0 + \beta_3 )
\sqrt{(\gamma'_{\rm max})^2-1}
-\frac{\gamma'_{\rm max} \mp \hat{\beta}_0 \sqrt{(\gamma'_{\rm max})^2-1}}
{(\gamma'_{\rm max})^2+\hat{\gamma}_0^2\hat{\beta}_0^2} \varpi
\right ]  ,
\label{einfhyd}
\end{eqnarray}
where $\gamma'_{\rm max}$ is given by equation (\ref{lofmaxrec}).
The plus and minus signs of the subscript in $\epsilon^\infty_\pm$
express the cases with the plasma velocity $v^{3\prime}=v'_{\rm max}$
and $v^{3\prime}=-v'_{\rm max}$, respectively.
If $\epsilon^\infty_- < 0$,
the energy-at-infinity of the plasma becomes negative.
Here we neglect the contribution from the electromagnetic field, because
when the plasma velocity becomes $v'_{\rm max}$,
most of the magnetic energy is converted to the plasma kinetic energy so that 
the electromagnetic energy-at-infinity becomes negligible.
Then the total energy-at-infinity of the plasma and the electromagnetic field
becomes negative. 
Here $\epsilon^\infty_-$ decreases monotonically when $u_{\rm A}$ increases,
and it is positive when $u_{\rm A}=0$ and negative when $u_{\rm A}$ is 
large enough in the ergosphere.
Then, we can define the one-valued function 
$U^-_{\rm A}(\hat{\beta}_0,\varpi,\alpha,\beta_3)$ which satisfies 
$\epsilon^\infty_- (U^-_{\rm A},\hat{\beta}_0,\varpi,\alpha,\beta_3)=0$.
The condition for the formation of the negative energy-at-infinity
through the magnetic reconnection is given by $u_{\rm A}>U^-_{\rm A}$.

The condition of escaping to infinity of the plasma particle
accelerated by the magnetic reconnection is given by
\begin{equation}
\Delta \epsilon^\infty_+ \equiv \epsilon^\infty_+ 
- \left ( 1-\frac{\Gamma}{\Gamma-1} \varpi \right ) > 0  .
\end{equation}
Here we assume the magnetic field is so weak far from the reconnection region
that we can neglect the interaction between the large-scale magnetic field
and the outflow from the reconnection region.
The monotonic function $\Delta \epsilon^\infty_+$
with respect to $u_{\rm A}$ is negative when $u_{\rm A}=0$
and positive when $u_{\rm A}$ is large enough.
Then we can also define the one-valued critical function
$U^+_{\rm A}(\hat{\beta}_0,\varpi,\alpha,\beta_3)$ such that 
$\Delta \epsilon^\infty_+ (U_{\rm A}^+,\hat{\beta}_0,\varpi,\alpha,\beta_3)=0$.
The condition of the escape to infinity of the plasma flow 
is written by $u_{\rm A}>U^+_{\rm A}$.

\begin{figure}
\epsscale{.8}
\plotone{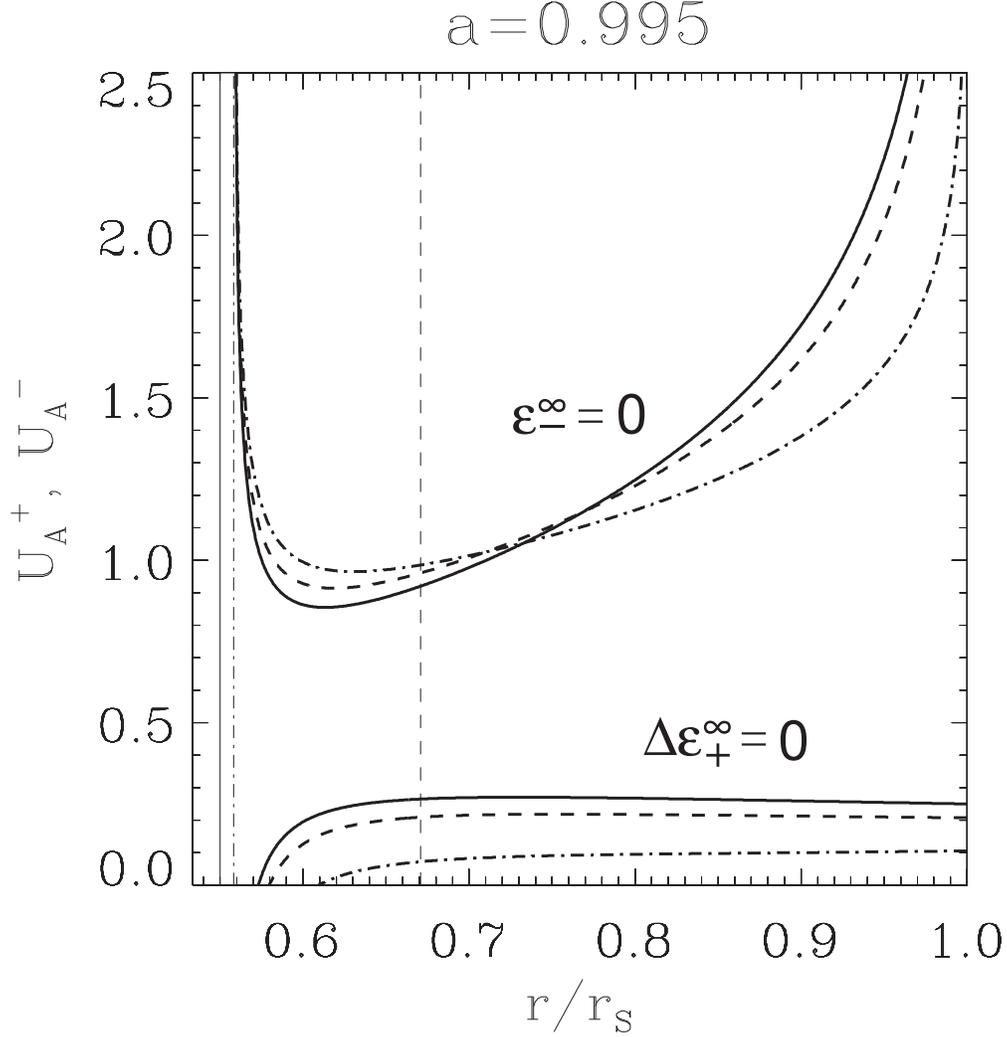}
\caption{
The critical Alfven four-velocity $U_{\rm A}^\pm$ for the energy extraction 
from the rotating black hole 
with the rotation parameter $a=0.995$
induced by the magnetic reconnection.
The base plasma rotates around the black hole with the Kepler
velocity $c\hat{\beta}_0=\hat{v}_{\rm K}$.
The thick solid, dashed, and dot-dashed lines correspond to the cases
of $\varpi=0, 0.2, 0.5$, respectively.
The lines with the sign $\Delta \epsilon^\infty_+=0$ ($\epsilon^\infty_-=0$) 
show the critical Alfven four-velocity
of the condition on escaping to infinity of the plasma accelerated by the 
magnetic reconnection (the negative energy-at-infinity plasma formation).
The vertical thin solid line at $r_{\rm H}=0.550 r_{\rm S}$
shows the black hole horizon. The vertical thin dashed line
at $r_{\rm ms} = 0.671 r_{\rm S}$ indicates the radius of the marginal
stable orbit. The vertical thin dot-dashed line at 
$r_{\rm L} = 0.559 r_{\rm S}$ shows the point where the Kepler
velocity is the light velocity $c$. Between the vertical thin solid 
line and the vertical thin dot-dashed line, there is no circular
orbit of a particle.
Generally speaking, $U^-_{\rm A}$ is infinite at $r=r_{\rm S}$,
while $U^+_{\rm A}$ is finite at this point.
\label{ua0995}}
\end{figure}

We apply the conditions $u_{\rm A} > U^+_{\rm A}$ and 
$u_{\rm A} > U^-_{\rm A}$ to the plasma with the rotation bulk velocity
$\hat{v}^3=c\hat{\beta}_0=\hat{v}_{\rm K}$.
%
Figure \ref{ua0995} shows $U^-_{\rm A}$ and $U^+_{\rm A}$ 
against $r/r_{\rm S}$ 
($r_{\rm S} = 2 r_{\rm g}$) for the fixed black hole 
rotation parameter $a=0.995$ and the pressure parameter $\varpi = 0, 0.2, 0.5$.
The lines of the critical values of Alfven four-velocities $U^-_{\rm A}$ 
and $U^+_{\rm A}$ are drawn as the contours of 
$\epsilon^\infty_-=0$ and $\Delta \epsilon^\infty_+=0$, respectively.
The vertical thin dashed line indicates the inner most stable orbit
radius of a single particle. The vertical thin dot-dashed line
shows the point of $\hat{v}_{\rm K} =c$. Then, the Kepler motion is unstable
between the dot-dashed line and the dashed line.
In the left region of the dot-dashed line, there is no
circular orbit anymore. The vertical thin solid line indicates the horizon
of the black hole. The upper regions of the thick lines $\epsilon^\infty_-=0$ 
show the condition of the formation
of the negative energy-at-infinity through the magnetic reconnection,
$u_{\rm A}=B_0/\sqrt{h_0}>U^-_{\rm A}$.
The different styles of lines indicate the different pressure cases
(solid line: $\varpi=0$, dashed line: $\varpi=0.2$, 
dot-dashed line: $\varpi=0.5$).
The regions above the thick lines $\Delta \epsilon^\infty_+ =0$ 
show the condition on escaping of the plasma
accelerated by the magnetic reconnection, 
$u_{\rm A} = B_0/\sqrt{h_0} > U^+_{\rm A}$.
The difference of the line styles denotes the same as that of the upper lines.
It is shown that the condition of the energy extraction
from the black hole through the magnetic reconnection is determined from the
condition of the formation of the negative energy-at-infinity,
$u_{\rm A}> U^-_{\rm A}$ in the case of the rotating plasma with the Kepler velocity.
In the zero pressure case ($\varpi=0$), the easiest condition is found at $r=0.61r_{\rm S}$, 
$u_{\rm A} \geq U^-_{\rm A}=0.86$. This means the relativistic reconnection
is required for the energy extraction from the black hole.
In the finite pressure case ($\varpi=0.2$), the condition is relatively relaxed around the outer
region of the ergosphere, while the condition around the inner ergosphere
is severer. However, the difference is small between these cases.
This case also requires the relativistic magnetic reconnection to
extract the black hole energy.
The condition of the case $\varpi=0.5$ is almost similar to the tendency of 
the previous two cases ($\varpi=0, 0.2$).

\begin{figure}
\epsscale{.8}
\plotone{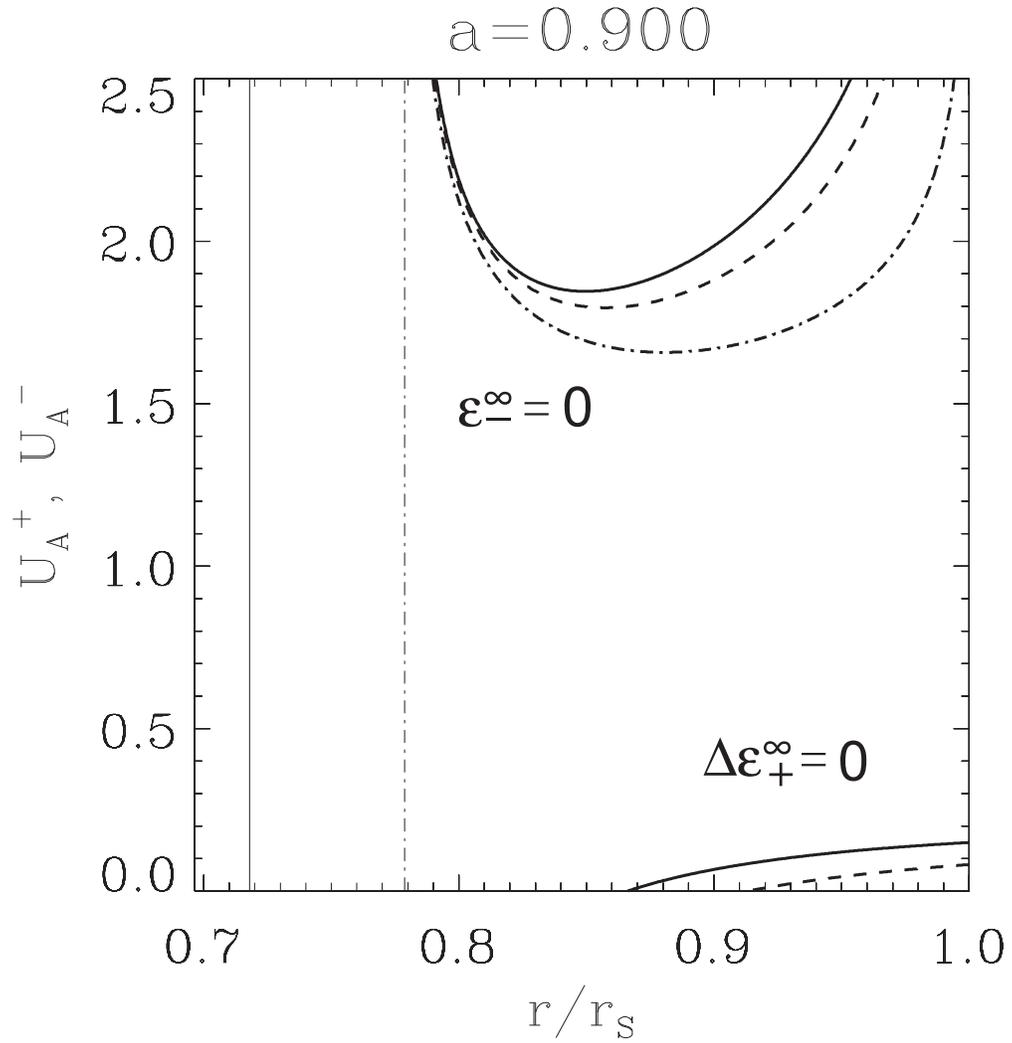}
\caption{
Similar to Fig. \ref{ua0995}, but for the black hole with the rotation
parameter $a=0.9$. The critical radii are $r_{\rm H} = 0.718 r_{\rm S}$, 
$r_{\rm L} = 0.779 r_{\rm S}$, and $r_{\rm ms} > r_{\rm S}$.
\label{ua0900}}
\end{figure}

Figure \ref{ua0900} shows the critical Alfven four-velocity
$U^\pm_{\rm A}$ of the energy extraction from the black hole with
the rotation parameter $a=0.9$.
In this case, the condition of the energy extraction from the black hole
becomes severer compared to the cases with the larger rotation 
parameters $a=0.995$. 

It is noted that in the case of the black hole rotation parameter
$a<1/\sqrt{2}$, there is no circular orbit inside of the ergosphere.
In such case, we can not consider the energy extraction through 
the magnetic reconnection in the circularly rotating plasma
around the black hole except for the case 
with the support by magnetic field.

\begin{figure}
\epsscale{.8}
\plotone{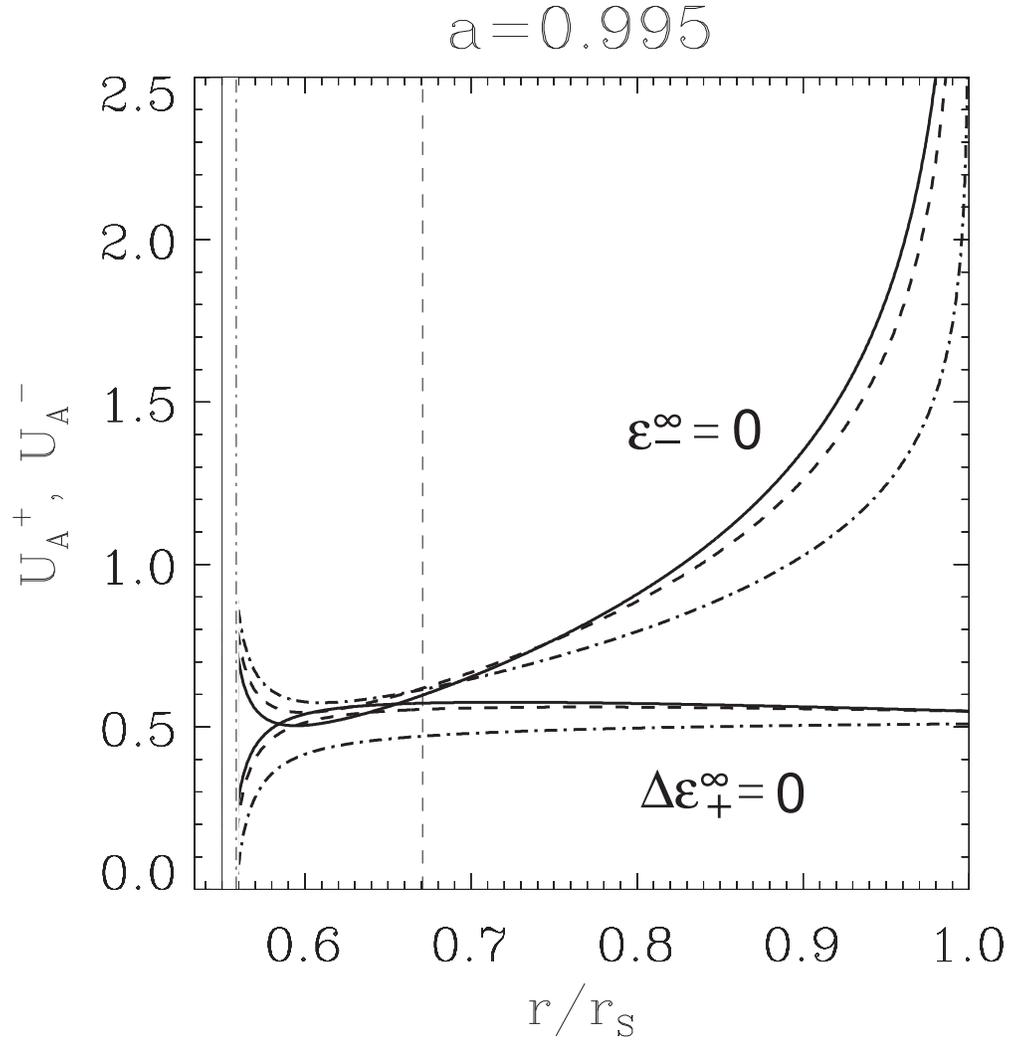}
\caption{
Similar to Fig. \ref{ua0995}, but for the sub-Keplerian case 
$c\hat{\beta}_0=0.6 \hat{v}_{\rm K}$.
\label{ua0995f06}}
\end{figure}

When we consider the slower rotating plasma case 
$c\hat{\beta}_0< \hat{v}_{\rm K}$, which may be an artificial assumption, 
the conditions
of the negative energy-at-infinity and the escape plasma become
comparable and relaxed as a whole.
Figure \ref{ua0995f06} shows the critical Alfven four-velocity of the case
$c\hat{\beta}_0 = 0.6 \hat{v}_{\rm K}$, $a=0.995$.
At $r=0.6r_{\rm S}$, $U^+_{\rm A} \simeq U^-_{\rm A}
\simeq 0.5$ for both the $\varpi =0$ and $\varpi =0.2$ cases.
This means the sub-relativistic magnetic reconnection can
extract the black hole energy.


\section{Discussion} \label{discussion}

In the previous section, we showed the possibility 
of the energy extraction from the black hole through the magnetic reconnection
in the ergosphere.
In this mechanism, the magnetic tension plays a significant role
to cause the plasma flow with the negative energy-at-infinity,
like the MHD Penrose process. If we consider quick magnetic reconnection,
the mechanism by the magnetic reconnection is more effective
than the MHD Penrose process,
because the fast plasma flow can be induced,
so that all magnetic energy can be converted to the kinetic 
energy of the plasma flow through the magnetic reconnection.

As the magnetic reconnection mechanism, we utilized
a rather simple model with an artificial resistivity
at the reconnection region and the local
approximation around the rotating black hole.
We further assumed that the outflow caused by
the magnetic reconnection is parallel to the azimuthal direction.
In general, the outflow is oblique to the azimuthal direction.
In the oblique case with the angle $\chi$ between the outflow and 
azimuthal directions, the condition of the energy extraction
through the magnetic reconnection is given by 
$u_{\rm A} > U^-_{\rm A}(u_{\rm A}, \varpi, \alpha, \beta_3 \cos \chi)$ 
and $u_{\rm A} > U^+_{\rm A}(u_{\rm A}, \varpi, \alpha, \beta_3 \cos \chi)$,
where $\beta_3$ of equation (\ref{einfhyd}) is replaced by $\beta_3 \cos \chi$.
This is a severer condition compared to the parallel case.

Let us briefly estimate the critical magnetic field required for the
energy extraction from the black hole through the magnetic reconnection
with respect to the AGNs, $\mu$QSOs, and GRBs.
Here the critical magnetic field in the SI unit, $B_{\rm crit}$,
is given by $B_{\rm crit}=\sqrt{\mu_0 h_0} \sim \sqrt{\mu_0 \rho_0}c$,
where $h_0$ and $\rho_0$ are the typical enthalpy and mass density
around the objects, respectively. Here, we assume the pressure is not
larger than $\rho_0 c^2$ and neglect it to get rough estimation.
To estimate the typical mass density $\rho_0$ of the plasma 
around the central black hole of these objects, we use
\begin{equation}
\rho_0 \simeq 5 \times 10^{4} 
\left ( \frac{\dot{M}}{M_\sun \, {\rm yr}^{-1}} \right )
\left ( \frac{M}{M_\sun} \right )^{-2}
\left (\frac{2r}{r_{\rm S}} \right )^{-3/2} \hspace{0.5cm} {\rm g \, cm^{-3}} ,
\label{estirho}
\end{equation}
where $\dot{M}$ is the accretion rate and $M$ is the black hole mass
\cite{rees84}.
For the AGN in the large elliptical galaxy M87, when
we assume $\dot{M}=10^{-2} M_\sun \,{\rm yr}^{-1}$, 
$r=r_{\rm S}$, and $M=3 \times 10^9 M_\sun$ 
\cite{reynolds96,ho99}, equation (\ref{estirho}) yields
$\rho_0 \simeq 2 \times 10^{-17} {\rm g \, cm}^{-3}$.
Then, we get $B_{\rm crit} \simeq 500$ G. This magnetic field is probable 
around a black hole of an AGN, thus the magnetic extraction of 
black hole energy is possible.

In a collapsar model with $\dot{M}=0.1 M_\sun {\rm s}^{-1}$ and
$M=3 M_\sun$ \cite{macfadyen99}, when we apply equation
(\ref{estirho}), we get 
$\rho_0 \simeq 6 \times 10^{9} {\rm g \, cm}^{-3}$ at $r=r_{\rm S}$
as mass density around a black hole in a GRB progenitor.
The critical magnetic field is then $B_{\rm crit} \simeq 8 \times 10^{15}$ G.
The magnetic field of the progenitors is estimated to be 
$10^{15}$G to $10^{17}$G \cite{putten99} and then
extraction of the black hole energy through the magnetic
reconnection is marginally probable in a core of a collapsar.

With respect to $\mu$QSO, GRS1915+105 has a mass accretion
rate of $\dot{M} = 7 \times 10^{-7} M_\sun \, {\rm yr}^{-1}$
\cite{mirabel94,fender04} with a mass
of $M=14M_\sun$ \cite{greiner01}.
Then equation (\ref{estirho}) yields 
$\rho_0 \simeq 6 \times 10^{-5} {\rm g \, cm}^{-3}$.
The critical magnetic field is estimated as $B_{\rm crit} \simeq 8 \times 10^8$ G.
This magnetic field is too strong as the field
around a black hole in $\mu$QSOs.
Thus, the energy extraction from a black hole in $\mu$QSO
through the magnetic reconnection may not be prospective.

\begin{figure}
\epsscale{.7}
\plotone{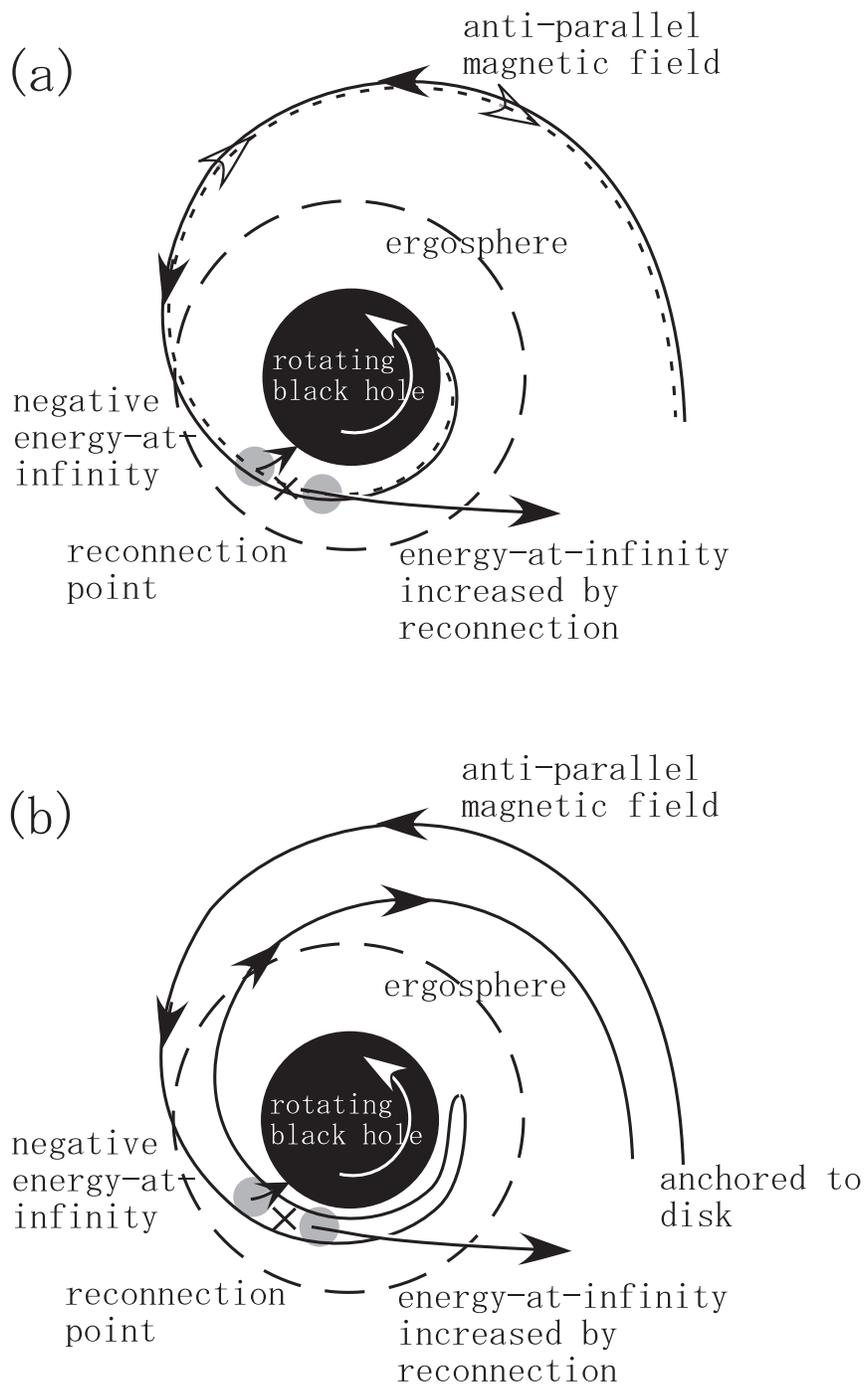}
\caption{
Astrophysical magnetic configuration of the magnetic reconnection
in the black hole ergosphere. (a) Anti-parallel magnetic field caused from 
the initial uniform magnetic field around the rapidly rotating black hole.
The solid/dashed line shows the anti-parallel magnetic field line
in front of/behind the equatorial plane from the reader.
In this magnetic field configuration, the current sheet locates
at the equatorial plane.
(b) Anti-parallel magnetic field formed by the closed magnetic
flux tube tied to the disk around the rapidly rotating black hole.
The inner part of the magnetic flux tube falls
into the black hole to elongate the flux tube,
and the anti-parallel magnetic field configuration is formed.
\label{skmpic}}
\end{figure}

We discuss the possibility of formation of anti-parallel
magnetic field with a current sheet where strong
magnetic reconnection in the ergosphere is caused.
First, let us consider uniform magnetic field around
a rotating black hole as the initial condition.
In this magnetic configuration, one may think that the magnetic reconnection
scarcely happens. However, this is caused naturally by the gravitation and
the frame-dragging effect of the rapidly rotating black hole.
Under this situation, GRMHD simulations were carried out with zero electric
resistivity \cite{koide02,koide03,komissarov04}.
The numerical simulations showed that the magnetic flux tubes across the
ergosphere are twisted by the frame-dragging effect of the rotating
black hole, and the plasma falling into the black hole makes the
magnetic field radial around the ergosphere. 
Here it is noted that the magnetic field line twisted by the frame-dragging
effect makes the angular momentum of the plasma around the equatorial
plane and the ergosphere negative ($l<0$), and the plasma with the
negative angular momentum falls into the black hole more rapidly.
The attractive force toward the black hole comes from 
the shear of the frame-dragging effect directly
(e.g. see the term with $\sigma_{ji}$ in equation (56) of Koide (2003)).
Beside the equatorial plane in the ergosphere, the magnetic field becomes
anti-parallel. The magnetic flux tubes are twisted strongly enough, and 
then the strong anti-parallel open magnetic field is formed almost 
along the azimuthal direction (Fig. \ref{skmpic}a).
In this way, the magnetic reconnection happens and the energy
of the rotating black hole is extracted through the magnetic reconnection,
even in the case of the initially uniform magnetic field.
When the magnetic reconnection happens around the anti-parallel magnetic field,
the outward flow from the reconnection region will be ejected toward
infinity along the open magnetic field lines. This outflow will be bent 
and pinched by the magnetic field and may become a jet.

Next, as the initial condition, we assume
closed magnetic flux tubes which are believed
to be formed in the accretion disks around 
the black holes \cite{putten99,koide06,mckinney06} (Fig. \ref{skmpic}b). 
When a single closed magnetic flux tube is tied to
an edge of a rotating quasi-stationary disk and a bulk part of the disk,
the plasma at the edge loses the angular momentum and falls into
the black hole, while the bulk plasma tied to the magnetic flux
tube increases the angular momentum and shifts outwardly.
The plasma at the disk edge falls spirally due to the frame-dragging 
effect, and the magnetic flux tubes dragged
by the plasma are elongated spirally.
If the twist of the magnetic flux tube is strong enough,
anti-parallel closed magnetic field is formed almost along the azimuthal direction. 
In this magnetic configuration, energy may be extracted from the black hole 
through the magnetic reconnection in the ergosphere.
The outflow from the reconnection region
will elongate the closed magnetic field lines, while this decelerates the outflow.
If the magnetic reconnection is caused in the elongated magnetic field lines,
the plasmoid is formed and is ejected to infinity.

As shown in the above two cases of the open and closed magnetic field, 
the large-scale dynamics of the outflow
through the magnetic reconnection depends on the large-scale magnetic field
configuration.
Here we note that closed magnetic flux tubes across an accretion disk
and an ergosphere around a rapidly rotating black hole is unstable and
expands vertically to form open magnetic field  
\cite{koide06,mckinney06}. Then, around a rapidly rotating black hole,
open magnetic field may be probable compared to closed field.
Anyway, these phenomena should be investigated by numerical
simulations of the full GRMHD with non-zero electric
resistivity (resistive GRMHD).

To be more exact, in both cases of the open and closed topologies 
of the magnetic field,
the outflow caused by the magnetic reconnection is influenced by
the large-scale magnetic field, and it is not determined only from the
local approximation which we used here.
In both cases, it is noted that the plasma with the negative
energy-at-infinity is farther 
from the horizon than the plasma accelerated by the magnetic reconnection.
If we assume the azimuthal symmetry of the initial condition,
interchange instability should be caused so that
the plasma with negative energy-at-infinity falls into the black hole
and the plasma with additional energy-at-infinity runs away to infinity.
With respect to the interchange instability, we consider essentially 
hydrodynamic mode where initially super-Keplerian inner part of the disk
supports sub-Keplerian outer part against the black hole gravity.
The closed magnetic flux tube is formed across the plasma with additional
energy-at-infinity. The plasma at the inner edge of the
magnetic loop falls into the black hole because of the deceleration 
by the magnetic tension, while the plasma at the outer edge of the
magnetic loop is accelerated and escapes to infinity.
Then the magnetic flux tube is elongated
between the escaping and falling plasmas. In such magnetic flux
tube, the anti-parallel magnetic field with strong current sheet
may be formed and the magnetic reconnection may be caused once again.
On the other hand, the plasma with the negative energy-at-infinity 
through the magnetic reconnection falls into the black hole 
and the magnetic flux tube tied to the plasma is also elongated 
by the frame-dragging effect.
The anti-parallel magnetic field in the magnetic flux tube will
also form and the magnetic reconnection is caused repeatedly.
Above discussion shows that the magnetic reconnection 
can be caused intermittently in the ergosphere.
To investigate these phenomena, the numerical simulations of
resistive GRMHD are also demanded.

The resistive GRMHD should solve the problems with respect to
the energy extraction through the magnetic reconnection.
For example, using resistive GRMHD simulations, 
we can take into account of the situation that 
the initial plasma falls into the black hole
with sub-Keplerian velocity of the plasma rotation,
which is neglected in this paper.
Unfortunately, no simulation with resistive GRMHD has been performed until now, while
recently, ideal GRMHD numerical simulations, where the electric
resistivity is zero, have come spread
(Koide et al. 1998, 1999, 2000, 2002, 2006; Koide 2003, 2004;
McKinney 2006; Punsly 2006; Komissarov et al. 2007, and references therein).
They confirmed important, interesting magnetic phenomena around the rotating
black holes, such as magnetically-induced energy extraction from
the rotating black hole
\cite{koide02,koide03,komissarov04} and formation of magnetically-driven 
relativistic jets \cite{mckinney06}.
To confirm that the plasma with additional energy 
by the magnetic reconnection
escapes to infinity and the plasma with the negative 
energy-at-infinity falls into the black hole, 
we may also use the ideal GRMHD simulations.
In such calculations, we can trace the plasma trajectories after the magnetic 
reconnection stops. The magnetic configuration of the post stage of the
magnetic reconnection is used as an initial condition.

On the other hand, in spite of the restriction of the ideal GRMHD,
many magnetic islands are seen in the last stages of long-term calculations 
\cite{mckinney04,koide06,mckinney06}.
These magnetic islands, of course, are artificial appearance.
However, these numerical results indicate that the magnetic
configuration where the magnetic reconnection occurs
is relatively easily formed around the black hole.
Recent X-ray observations of the solar corona confirmed
that the magnetic reconnection takes places frequently in
the active region of the corona and causes drastic phenomena,
like solar flares. The observations and the recent theories of solar 
and stellar flares indicate that the magnetic reconnection
is common in the astrophysical plasmas around the Sun, stars, and
the compact objects including black holes
\cite{masuda94,shibata97}.
The energy extraction from the black hole through the magnetic reconnection
is one of the phenomena of magnetic 
reconnection around the black hole. More drastic phenomena related
with the magnetic reconnection may exist. To investigate these
phenomena, resistive GRMHD numerical calculations will play 
a crucial role.

\acknowledgments

We thank Mika Koide, Takahiro Kudoh, Kazunari Shibata, and Masaaki Takahashi
for their help for this study.
We also thank Naoyuki Watanabe and Takaaki Yokoyama
for their permission of the use of data in their paper 
\cite{watanabe06}.
This work was supported in part by the Scientific Research
Fund of the Japanese Ministry of Education, Culture, Sports,
Science and Technology.






\appendix

\section{Relativistic adiabatic incompressible ball approach}

To take account of inertia effect of pressure into plasma, 
we use an approximation of incompressible fluid, which is consisted
of small, separated, constant volume elements.
The element is covered by thin, light, adiabatic, closed skin 
and its volume is constant, like a ball used for soft tennis. 
We call this method ``relativistic adiabatic incompressible ball (RAIB) approach".
Here we assumed gas pressure does not
work to the plasma and influences the plasma dynamics only
through an inertia effect.
Let us consider the fluid in one ball with the mass $m$.
When the ball locates at $\VEC{r}=\VEC{r}(t)$, the mass density
of the gas is
\begin{equation}
\rho(\VEC{r},t) = \frac{m}{\gamma(t)} \delta^3(\VEC{r}-\VEC{r}(t))  ,
\label{apmasden}
\end{equation}
where $\gamma(t)$ is the Lorentz factor of the ball at time $t$ and
$\delta^3(\VEC{r})$ is the Dirac's $\delta$-function in
three-dimensional space.
Because the gas in the ball is assumed to be incompressible and adiabatic 
and then its temperature is constant,
the pressure should be proportional to the mass density,
$p(\VEC{r},t) \propto \rho(\VEC{r},t)$. 
On the other hand, the thermal energy in the ball
\begin{equation}
U = \int_\Omega \frac{p(\VEC{r},t)}{\Gamma-1} \gamma(t) dV
\end{equation}
should be constant. Then we found
\begin{equation}
p(\VEC{r},t) = \frac{(\Gamma-1) U}{\gamma(t)} 
\delta^3(\VEC{r}-\VEC{r}(t))   .
\label{apppres}
\end{equation}
Using equations (\ref{eatinfhyd}), (\ref{apmasden}), and (\ref{apppres}), 
we found the energy-at-infinity of the gas in the ball as,
\begin{eqnarray}
E^\infty &=& 
\int_\Omega
\left( \hat{e}_{\rm hyd} + \sum_i c\beta_i \frac{\mathfrak h}{c^2} 
\hat{\gamma}^2 \hat{v}^i \right) dV  \nonumber \\
         & = & \int_\Omega \alpha \left [ 
\left( \rho c^2 +\frac{\Gamma}{\Gamma -1}p \right) 
\hat{\gamma}^2 (1 + \beta_3 \hat{\beta}^3) -p  \right ] dV \nonumber \\
         & = & \alpha \left [ \left ( \hat{\gamma} + \beta_3 \hat{U}^3 \right ) H
            - \frac{\Gamma-1}{\hat{\gamma}} U \right ],
\label{eatinficstba}
\end{eqnarray}
where $\hat{\beta}^3=\hat{v}^3/c$ and $H=mc^2 + \Gamma U$.
In the special relativistic case, equation (\ref{eatinficstba})
yields the total energy of the ball as,
\begin{equation}
E_{\rm tot} = \gamma H - \frac{\Gamma-1}{\gamma} U   .
\label{apptotengspe}
\end{equation}

Next we derive the energy-at-infinity of the incompressible ball rotating
circularly around the black hole.
We assume that the bulk plasma rotates circularly with the velocity
$\hat{v}^3=c\hat{\beta}_0$, and then the relative three-velocity
between the rest frame of the bulk plasma and the ZAMO frame is
$\hat{v}^3=c\hat{\beta}_0$, $\hat{v}^1=\hat{v}^2=0$.
The line elements of the ZAMO frame $(c\hat{t},\hat{x}^1,\hat{x}^2,\hat{x}^3)$ and 
the bulk plasma rest frame $(ct',x^{1\prime},x^{2\prime},x^{3\prime})$ 
are related by the Lorentz transformation.
Using equations (\ref{eatinficstba}) and the Lorentz transformation, 
the energy-at-infinity of the incompressible ball with four-velocity 
observed by the bulk plasma rest frame 
$(\gamma',U'^1, U'^2, U'^3)$ is given by
\begin{eqnarray}
E^\infty 
 & = & \alpha H \hat{\gamma}_0 \left [
\left ( 1 + \beta_3 \hat{\beta}_0 \right ) \gamma'
+ \left ( \hat{\beta}_0 + \beta_3 \right ) U'^3
- \frac{\gamma'-\hat{\beta}_0 U'^3}{\gamma'^2+\hat{\gamma}_0^2 \hat{\beta}_0^3} 
\varpi \right ]  .
\label{appengatinfbal}
\end{eqnarray}
It is noted that the four-velocity of the plasma is related with 
the Lorentz factor by $U'^3=\pm \sqrt{\gamma'^2-1}$ in the case of
$U'^1=U'^2=0$. This formula of the energy-at-infinity of the
incompressible ball (\ref{appengatinfbal}) is applicable to that of 
one particle of the plasma effectively.

\clearpage

\clearpage


\end{document}